\documentclass[useAMS,usegraphicx,usenatbib]{mn2e}

\def\ion#1#2{#1$\,${\uppercase\expandafter{\romannumeral1}}\relax}

\title[\ion{H}{1} in NGC 5433 and its Environment]{\ion{H}{1} in NGC~5433 and its Environment: High-Latitude Emission in a Small Galaxy Group}
\author[Spekkens, Irwin and Saikia]
       {Kristine Spekkens$^1$, Judith A. Irwin$^2$, and D. J. Saikia$^3$
\\
   $1$ Cornell University, Dept. of Astronomy, 516
        Space Sciences Building, Ithaca, 
         NY 14853-6801\\
   $2$ Dept. of Physics, Queen's University, Kingston, Canada, K7L 3N6\\
   $3$ National Centre for Radio Astrophysics, Tata Institute of
Fundamental Research, Pune University Campus, Post Bag 3, \\Pune 411 007, India
}
\begin{document}
\date{Accepted.    Received }

\pagerange{\pageref{firstpage}--\pageref{lastpage}}
\pubyear{2004}


\maketitle

\label{firstpage}

\begin{abstract}
We present \ion{H}{1} synthesis maps of the edge-on starburst NGC~5433 and its environment, obtained with the VLA in its C and D configurations. 
The observations and spectral model residuals of the main disc emission in NGC~5433 reveal 3 extraplanar features. We associate 2 of these features with coherent extraplanar extensions across multiple spectral channels in our data, including a complete loop in position-velocity space. Interpreting the latter as an expanding shell we derive a corresponding input energy of $2 \times 10^{54}$ ergs, comparable to that for the largest supershells found in the Galaxy and those in other edge-on systems. 
NGC~5433 is in a richer environment than previously thought. We confirm that KUG~1359+326 is a physical companion to NGC~5433 and find two new faint companions, both with Minnesota Automated Plate Scanner identifications, that we label SIS-1 and SIS-2. Including the more distant IC~4357, NGC~5433 is the dominant member of a group of at least 5 galaxies, spanning over 750~kpc in a filamentary structure. A variety of evidence suggests that interactions are occurring in this group.  While a number of underlying mechanisms are consistent with the morphology of the high-latitude features in NGC~5433, we argue that environmental effects may play a role in their generation.


\end{abstract}


\section{Introduction}
High-latitude extensions from the interstellar medium (ISM) in spiral galaxies are important probes of disc stability and energetics, halo dynamics, and the physical processes that drive galaxy evolution.  
These features are best detected in edge-on
galaxies, in which extraplanar emission can be clearly separated from the underlying
disc.  Discrete features (e.g. arcs, plumes, and filaments in various
ISM tracers, or `supershells' in \ion{H}{1} as well as
 broader-scale structures such as thick discs or
haloes) have now been identified in many galaxies using this approach.
This has led to 
ideas of galactic circulation involving `fountains', `chimneys',
or possibly outflowing
winds which may enrich the intergalactic medium 
(Shapiro and Field 1976, 
Bregman 1980, Norman and Ikeuchi 1989, Heckman, Armus \& Miley 1990,
Breitschwerdt, V{\"o}lk \& McKenzie 1991, Breitschwerdt, McKenzie \& V{\"o}lk 1993).

Disc-halo features are usually
attributed to underlying drivers such as supernovae and stellar winds.
However, these processes alone may be insufficient to produce the highly energetic \ion{H}{1} supershells seen in some galaxies, or to account for the absence of detectable                  
star formation regions or remnants in others (e.g. Rhode et al. 1999, Perna \& Gaensler 2004).  
Thus, while star formation may be a key ingredient in the disc-halo phenomenon,
it may not, alone, be sufficient to drive disc-halo dynamics.
Impacting clouds
have long been invoked as a possible alternative (Tenorio-Tagle \& Bodenheimer 1988,
Santill\'{a}n et al. 1999), but galaxies
that are apparently isolated also sometimes display large supershells
(King \& Irwin 1997, Lee \& Irwin 1997).
This has led to speculation about other possible contributors, including
gamma-ray bursts (Efremov, Elmegreen \& Hodge 1998, Loeb \& Perna 1998), inflated bubbles from 
radio jets
(Gopal-Krishna \& Irwin 2000) and magnetic fields (Kamaya et al. 1996). 
Environmental effects have also been invoked to explain a variety of halo phenomena, such as differing extents of radio 
(Dahlem, Lisenfeld \& Golla 1995) and X-ray haloes (Wang, Chaves \& Irwin 2003) among various edge-on systems.

This paper represents an attempt to further characterize the importance of environment in driving disc-halo
dynamics.  A first problem is the lack of information about the environments of galaxies 
showing disc-halo outflows. 
We have therefore obtained maps of the large-scale \ion{H}{1} in the edge-on 
galaxy NGC~5433 using the Very Large Array\footnote{The VLA is a facility of the National Radio Astronomy Observatory (NRAO). The NRAO is a facility of the National Science Foundation operated under cooperative agreement by Associated Universities, Inc.} (VLA)
 in its C and D configurations.  We examine the main disc and extraplanar \ion{H}{1} in NGC~5433 using the higher resolution C configuration data, and combine the C and D datasets to probe its environment.

NGC~5433 ($D$ = 65 Mpc; $H_0 = 70\,\,\rm{km\,s^{-1}\,Mpc^{-1}}$)
is an infrared bright spiral (Soifer et al. 1989) 
that was observed in the radio continuum surveys of Irwin, English \&
Sorathia (1999; see their fig.~19) and Irwin, Saikia \& English (2000; see their fig.~1).
The 20 cm images from these studies reveal a thick disc which resolves into discrete disc-halo
features at high resolution.  
Single dish \ion{H}{1} observations of NGC~5433 have been obtained by 
Mirabel \& 
Sanders (1988) and Schneider et al. (1990) and a
low sensitivity VLA integrated \ion{H}{1}  map
has been presented by Thomas et al. 
(2002).

\begin{figure*}
\includegraphics{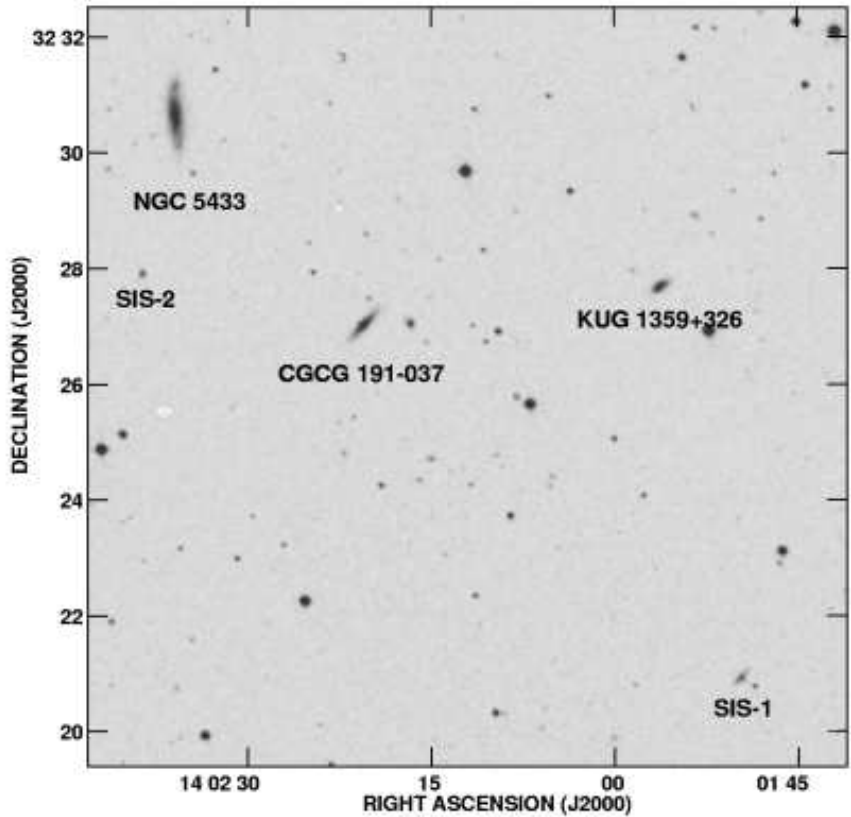}
\caption{DSS image of NGC 5433 and its SW environment. Labels have been placed immediately below the galaxy. The more distant galaxy IC~4357 is not shown; it is located 44 arcmin SW of NGC~5433, along a line joining NGC~5433 and CGCG~191-037.
\label{fig1}
}
\end{figure*}

An optical DSS image of NGC~5433 and its southwest (SW) environment is shown in
Fig.~\ref{fig1}.  NGC~5433 appears slightly curved, with an outer
ring which is distorted on the north side. Its RC3 (de Vaucouleurs et al. 
1991) classification as an Sdm has been revised to an earlier type by Dale et al. (2000), who include 
NGC~5433 in an {\it Infrared Space Observatory} Key Project study of normal star-forming galaxies (see also Malhotra et al. 2001). 
X-ray (Rephaeli, Gruber \& Persic 1995) and submillimetre (Thomas et al. 2002) 
detections have also been reported. The basic properties of NGC~5433, obtained from the NASA/IPAC Extragalactic Database (NED) unless
otherwise noted, are given in Table~\ref{basic}.

Two galaxies in the vicinity of NGC~5433, located 
5 arcmin (CGCG~191-037) and 9 arcmin (KUG~1359+326) to the SW, have been 
identified as possible companions in the 
Uppsala General Catalogue of Galaxies
(Nilson 1973). More recently, the trio have been 
listed as WBL~485 by White et al. (1999) in their
catalogue of nearby poor clusters of galaxies. 
 On larger scales NGC~5433 has been paired with IC~4357, 
44 arcmin to the SW and differing in recessional velocity from NGC~5433 by $40~\rm{km\,s^{-1}}$, in the Nearby 
Optical Galaxies (NOG) group assignment of Giuricin et al. (2000).
Several other faint galaxies,
identified using the Minnesota Automated Plate Scanner (MAPS)
by Cabanela (1999) or in the 2 Micron All-Sky Survey (2MASS), are also in the field shown in Fig.~\ref{fig1}. 
Two of these have been labelled SIS-1 (MAPS-NGP O\_325\_0013717)
and SIS-2 (MAPS-NGP O\_271\_0297230) for simplicity.  
There are no galaxies brighter than 17th (blue) magnitude in regions of the same size in other directions with respect to NGC~5433.
 Available data (from NED) for the galaxies labelled in Fig.~\ref{fig1} 
are given in Table~\ref{basic}.


In this paper we examine the \ion{H}{1} distribution and kinematics of NGC~5433, and characterize its environment for the first time. 
Our observing and data reduction techniques are given in \S\ref{obsred}, 
and details of the radio frequency interference (RFI) excision we performed are in Appendix~\ref{rfi}.   
In \S\ref{results} we present our results: an analysis of the \ion{H}{1} 
content and kinematics of NGC~5433 is in \S\ref{HI_n5433},
and in \S\ref{HI_environment}~and~\S\ref{HI_companions} we 
characterize the environment of NGC~5433 and the 
\ion{H}{1} in its companions. In \S\ref{discussion} we discuss the prevalence of high-latitude \ion{H}{1} in edge-on spirals, the characteristics of the NGC~5433 group and possible origins for the extraplanar emission observed. A summary of our findings is in \S\ref{conclusions}.


\begin{table*}
 \begin{minipage}{140mm}
  \caption{Basic Properties of NGC 5433 and Nearby Galaxies \label{basic}}
  \begin{tabular}{@{}lcccccccc@{}}
\hline
Galaxy & Morph. & $\alpha$ (J2000) & $\delta$ (J2000) & Sep.$^a$ & Mag.  & Diameter
& $V_{sys}$$^b$ & $i$$^c$ \\
& Type & ($^{\rm h}$ $^{\rm m}$ $^{\rm s}$)
& ($^\circ$ $\arcmin$ $\arcsec$)
& (arcmin)
& (blue)
& (blue; arcmin$^2$)
& ($\rm{km}\,\rm{s}^{-1}$)
& ($^\circ$) \\
(1) & (2) & (3)
& (4) & (5) & (6) & (7) & (8) & (9)\\ 
\hline
NGC~5433 & SAB(s)c:$^d$ & 14 02 36.05 & 32 30 37.8 & ... & 14.10 & $1.6\,\times\,0.4$ & $4354\,\pm\,7$ & 78
\\
CGCG 191-037 & ... & 14 02 20.56 & 32 26 53.5 & 5.0 & 15.6 & $0.8\,\times\,0.2$ & ... & 76\\
KUG~1359+326 & ...& 14 01 55.87 & 32 27 29.6 & 9.0 & 15.7 & $0.4\,\times\,0.2$ & ... & 60\\ 
SIS-1$^e$     & ...& 14 01 49.62 & 32 20 34.5 & 14.0 & 17.45 & $0.37\,\times\,0.17$ & ... & 63\\
SIS-2$^e$     & ...& 14 02 38.97 & 32 27 50.3 & 2.9 & 18.00 & $0.21\,\times\,0.16$ & ... & 40\\
IC~4357       & S? & 14 00 43.69 & 31 53 39.0 & 44.0 & 14.75 &  $1.1\,\times\,0.6$ & $4394\,\pm\,36$ & 60\\
\hline
\end{tabular}
$a)$ Projected angular separation from NGC~5433.\hfill\break
$b)$ Heliocentric, optical definition (and so throughout).\hfill\break
$c)$ From Col. 7 assuming a thin disc except for NGC~5433, where an intrinsic axial ratio $q=0.13$ is adopted.\hfill\break
$d)$ Revised from RC3 (de Vaucouleurs et al. 1991) classification of Sdm, by Dale et al. (2000).\hfill\break
$e)$ Alternates: SIS-1 = MAPS-NGP O\_325\_0013717, SIS-2 = MAPS-NGP O\_271\_0297230\hfill\break
\end{minipage}
\end{table*}

\section[]{Observations and Data Reduction}
\label{obsred}

The data were obtained with the VLA in its C and D configurations on 
2001 July 21, 23 and 1999 May 4, respectively.  NGC~5433 was at the pointing centre of each observation, and as such all of the galaxies listed in 
Table~\ref{basic} except IC~4357 fell within the instrument's primary beam. During each run the galaxy was observed in 25--35 minute 
intervals, separated by 3--5 minute phase calibrator observations. 
Online Hanning smoothing was applied. The flux calibrator, also used as a bandpass calibrator, was observed at both 
the start and end of each observing run.  A total of 554 minutes were spent on source in the C configuration, and 142 minutes in the D configuration. Windy conditions throughout the D configuration run caused a number of antennas to stow, reducing the amount of data obtained at those times by an average of 10 per cent. 

 The data from each observing run were calibrated separately using the Astronomical Image Processing System ({\sevensize AIPS}). Standard flux, phase and bandpass calibration routines were applied, and the continuum emission in the resulting cubes was removed by subtracting linear fits to the real and imaginary parts of the visibility data in the line-free frequency channels of each bandpass. Following these steps, strong RFI was detected in the central 1 or 2 channels of each dataset. The culprit was the 1400 MHz band centre (corresponding to the recessional velocity of NGC~5433) adopted for the observations: with this setup, correlated noise from the 7th harmonic of the VLA's 200 MHz local oscillator at L-band contaminated the shortest east-west baselines in the central channels, for which the expected fringe rate is zero (Bagri 1996). This RFI was excised by {\it a)} removing baselines with east-west projections close to the array centre {\em in all channels}, and {\it b)} clipping contaminated data in the visibility domain from {\em infected central channels only}. With this approach the RFI was excised at the root mean square (RMS) noise level $\sigma$ of the infected channels. The main results of the RFI removal are that $\sigma$ for the central channel of each dataset is $\sim$10 per cent higher than the corresponding datacube mean, and flux estimates for NGC~5433 from the central 2 channels of the C configuration data have an associated {\em systematic error} on the order of $\sigma$; all other flux measurements and HI maps are unaffected. Details of the RFI excision procedure are given in Appendix~\ref{rfi}.

After calibration, the C configuration datasets were combined in 
the visibility domain and imaged using a variety of 
weighting schemes to emphasize \ion{H}{1} structures of different spatial scales. 
The C and D configuration data were also combined and imaged in a 
similar manner; we will refer to this combination as C+D data. Each 
data cube was then {\sevensize CLEAN}ed (Clark 1980) and corrected for the attenuation 
of the primary beam.   Due to the lower sensitivity (relative to the 
C+D maps) and resolution (relative to the C maps) of the D configuration data, we show no images of this dataset alone. However, the C and D 
configuration observations were obtained independently and have similar 
sensitivities (see Table~\ref{observe}), and thus the D data provide 
a useful consistency check on the reality of faint features detected in 
the C data: one expects to find them in both cubes.  
All maps displayed 
are uncorrected for the primary beam, but calculations are performed 
on the corrected cubes. 
The observing and map parameters for the C, D and C+D data 
are given in Table~\ref{observe}. 

A map of the continuum emission was also made from the line-free channels 
and used to measure continuum fluxes in the vicinity of our \ion{H}{1} 
detections (Table~\ref{global_params}).  
Since the resulting images of NGC~5433 do
not improve upon those of Irwin et al. (1999), we do
 not reproduce them here.

\begin{table*}
 \begin{minipage}{125mm}
  \caption{\ion{H}{1} Observing and Map Parameters \label{observe}}
  \begin{tabular}{@{}lccc@{}}
\hline
Parameter &  C config. & D config. &  C+D combined \\ 
(1) & (2) & (3) & (4)\\
\hline
Observing date & 2001 July 21 \& 23 & 1999 May 4 & ---\\
On-source observing time $\,$ (min.) & 554 & 142 & 696\\
Band centre $\,$ (km s$^{-1}$) & 4352  & 4352 & 4352 \\
Total bandwidth $\,$(km s$^{-1}$) & 657 & 657 & 657\\ 
Channel width $\,$ (km s$^{-1}$) & 10.1  &  21.2 & 21.2 \\ 
Natural weighting: & & & \\
$\,\,\,\,$ Synth. beam $\,$ (arcsec) $\,$ @ PA $\,$ (\degr) & $19\,\times \, 17$ @ 70
 & $53\,\times\,50$ @ -24 & $30\, \times \, 29$ @ -54
\\
$\,\,\,\,$ RMS map noise $\sigma$$^a$ $\,$ (mJy $\mathrm{beam}^{-1}$) & 0.31 &  0.36 & 0.21
\\
Robust weighting$^b$: & & & \\
$\,\,\,\,$ Synth. beam $\,$ (arcsec) $\,$ @ PA $\,$ (\degr) & $16\times \, 15$ @ 74
 & $47\,\times\,45$ @ -24 & $23\, \times \, 21$ @ -58
\\
$\,\,\,\,$ RMS map noise $\sigma$$^a$ $\,$ (mJy $\mathrm{beam}^{-1}$) & 0.32 & 0.38 & 0.21
\\
Uniform weighting: & & & \\
$\,\,\,\,$ Synth. beam $\,$ (arcsec) $\,$ @ PA $\,$ (\degr) & $13\,\times \,13$ @ 68
 & $42\,\times\,39$ @ -26 & $16\, \times \,15$ @ -53
\\
$\,\,\,\,$ RMS map noise $\sigma$$^a$ $\,$ (mJy $\mathrm{beam}^{-1}$) & 0.37 & 0.41 & 0.24
\\
\hline
\end{tabular}
$a)$ $\sigma$ at $V=4352\,\,\rm{km\,s^{-1}}$ is $\sim$10 per cent higher than this value; see text and Appendix~\ref{rfi}.\hfill\break
$b)$ {\sevensize AIPS} robustness parameter is chosen to produce a synthesized beam intermediate to natural and uniform weighting.\hfill\break
\end{minipage}
\end{table*}

\begin{figure}
\includegraphics{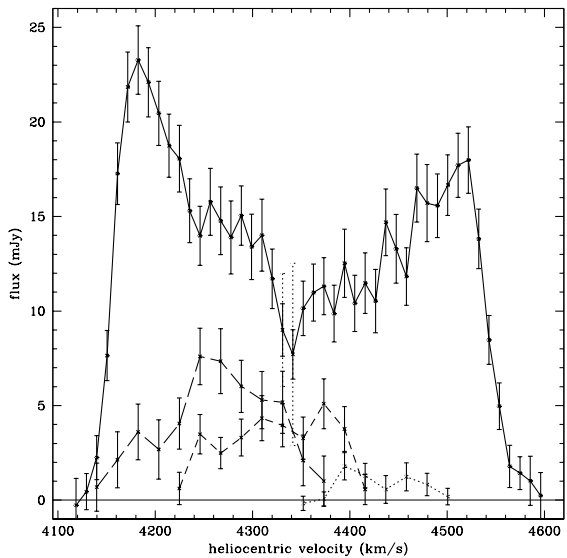}
\caption{
Global profiles for NGC~5433 {\it (solid lines)}, 
KUG 1359+326 {\it (long-dashed lines)}, SIS-1 {\it (short-dashed lines)}, and SIS-2 {\it (dotted lines)}.  Solid error bars represent 1$\sigma$ random errors. For NGC~5433, the dotted part of the error bars represents additional systematic errors on those points from RFI removal; see \S\ref{obsred} and Appendix~\ref{rfi} for details. 
\label{fig2}
}
\end{figure}


\section{Results}
\label{results}

 In this section we present our findings regarding the \ion{H}{1} content of NGC~5433 and its environment. In \S\ref{HI_n5433} we use the higher resolution C configuration data to examine in-disc and extraplanar \ion{H}{1} of NGC~5433, and adopt the higher sensitivity C+D data to characterize the environment of NGC~5433 in \S\ref{HI_environment}. The C+D data are also used to study the \ion{H}{1} content of the detected companions in \S\ref{HI_companions}.
Throughout, we show images of datacubes with weightings that highlight the features we wish to discuss.

\subsection{\ion{H}{1} in NGC~5433}
\label{HI_n5433}

The spatial/spectral resolution and sensitivity of our C configuration data allow us to examine the morphology and kinematics
of the \ion{H}{1} in NGC~5433 in some detail. We discuss its
global \ion{H}{1} content and parameters in \S\ref{global_n5433}, model
its \ion{H}{1} distribution in \S\ref{model}, 
and characterize its extraplanar \ion{H}{1} extensions in \S\ref{extensions}.

\subsubsection{Global \ion{H}{1} content}
\label{global_n5433}

The global profile of NGC~5433, obtained from the naturally weighted C configuration data, is shown in Fig.~\ref{fig2}.
Channel maps are given in Fig.~\ref{fig3a}:
 \ion{H}{1} is detected from  4586~km~s$^{-1}$ to 4151~km~s$^{-1}$, with the north side of the galaxy receding and the
south side advancing. 
 Both figures
show that
 \ion{H}{1} emission is stronger on the southern side of the galaxy.
 This asymmetry 
is also seen in the single-dish global profiles of Mirabel \& Sanders (1988) and
Schneider et al. (1990).  Note that the ``dip''
in the global profile near 4350~km~s$^{-1}$ may not be real, given the larger
(dotted) error bars in these channels (see \S\ref{obsred} and Appendix~\ref{rfi}).

\begin{figure*}
\includegraphics{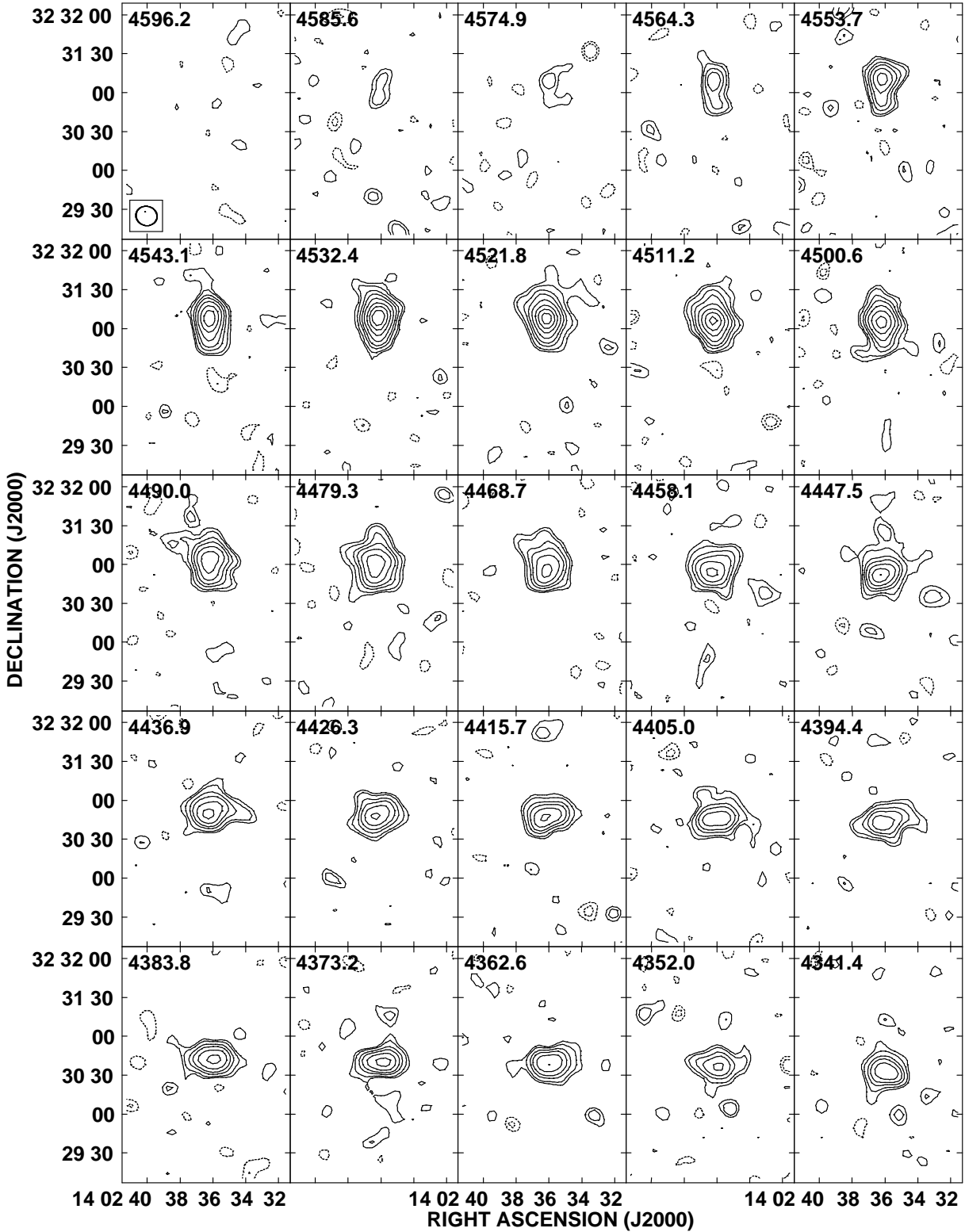}
\caption
{Robustly weighted C configuration channel maps for NGC 5433.  
Contours are at 0.31~$\,\times\,$~(-3 (dashed), -2 (dashed), 2 (2$\sigma$), 3, 5, 7, 10, 14, 18, 22, 26, 30, 32, 34, 36, 37, 37.5) $\rm{mJy}\,\rm{beam}^{-1}$. The channel velocity in $\rm{km\,s^{-1}}$ is indicated at the top of each frame, and the synthesized beam is given in the lower left corner of the first frame.
\label{fig3a}
}
\end{figure*}

\begin{figure*}
\includegraphics{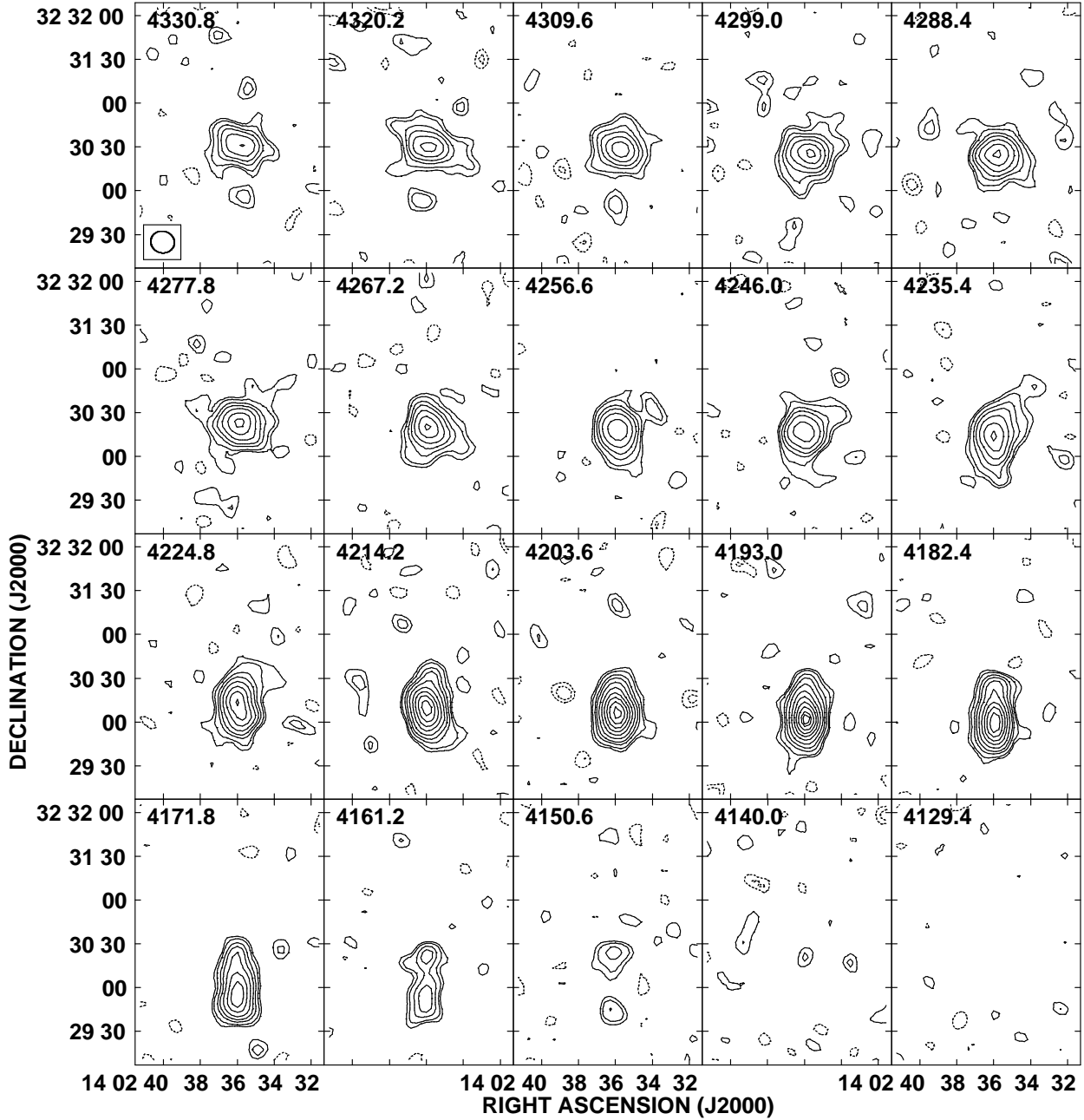}
\addtocounter{figure}{-1}
\caption
{Continued. 
\label{fig3b}
}
\end{figure*}

Fig.~\ref{fig4} shows the total intensity (zeroth moment)
and intensity-weighted velocity field (first moment) of NGC~5433
from uniformly weighted C configuration data.
A mild warp is present, most easily seen toward
the ends of the optical disc.  This warp is complex, showing
curvature toward the west (east) at the northern (southern)
edges of the optical disc and in the opposite
direction (east on the north side, west on the south side) farther out.
3 features extending from the disc, labelled F1 through F3, 
are also visible. From the mean column densities and sizes of the features in Fig.~\ref{fig4}a, we estimate that an \ion{H}{1} mass $M_{HI}\approx\,10^8\,M_{\odot}$ is associated with each.
The velocity field (Fig.~\ref{fig4}b) is well ordered, but the  
kinematic major and minor axes are not quite perpendicular.  This may indicate the presence of 
a bar or other oval distortion in the disc, and is perhaps expected given NGC~5433's morphological classification as a barred galaxy (Dale et al. 2000). 
The extensions F1 -- F3 show no kinematically distinct signatures in the first moment map.

A position-velocity plot for a 3 arcsec--wide slice from the uniformly weighted data along the major axis of NGC~5433 is given in Fig.~\ref{fig5}. 
It reveals that the disc exhibits largely solid-body rotation out to 
 20 arcsec (4 kpc, after correction for beam smearing) from the centre.  
The rotation curve of the galaxy flattens out only in the \ion{H}{1}-rich, 
southern side of the disc. From this figure as well as 
Fig.~\ref{fig4}a, it is clear that the northern \ion{H}{1} does not extend as far out as the southern gas.

\begin{table*}
 \centering
 \begin{minipage}{135mm}
  \caption{Global Properties of the Galaxies \label{global_params}}
  \begin{tabular}{@{}lcccccc@{}}
\hline
Galaxy & $V_{sys}$$^a$ & $W_o\sin{i}$$^b$ 
& ${\int}S\,dV$ & $M_{HI}$ & $M_T$ & $S_{20}$$^c$\\
& $(\rm{km}\,\rm{s}^{-1})$ &  $(\rm{km}\,\rm{s}^{-1})$ &  $(\rm{Jy}\,\rm{km}\,\rm{s}^{-1})$
& ($\,\times\,$~10$^9 M_{\odot}$) & ($\,\times\,$~10$^{11}~M_{\odot}$) & (mJy)\\
(1) & (2) & (3) & (4) & (5) & (6) & (7)\\
\hline
NGC~5433$^d$ & 4346 ${\pm}$ 7 & 381 ${\pm}$ 7 & 5.89 ${\pm}^{0.11r}_{0.04s}$ & 5.89 ${\pm}^{0.11r}_ {0.04s}$ & 1.64 ${\pm}$ 0.14 & 62.8 ${\pm}$ 3.2\\
KUG~1359+326 & 4282 ${\pm}$ 29 & 123 ${\pm}$ 29 & 1.01 ${\pm}$ 0.11 & 1.01 ${\pm}$ 0.11 & 0.10 ${\pm}$ 0.03 & 2.0 ${\pm}$ 0.4\\
SIS-1& 4321 ${\pm}$ 16 & 163 ${\pm}$ 16 & 0.66 ${\pm}$ 0.07 & 0.66 ${\pm}$ 0.07 & 0.16 ${\pm}$ 0.02 & $<0.59$ \\
SIS-2& 4431 ${\pm}$ 19 & 91 ${\pm}$ 19 & 0.12 ${\pm}$ 0.04 &0.12 ${\pm}$ 0.04 & 0.05 ${\pm}$ 0.01 & $<0.36$\\
\hline
\end{tabular}
$a)$ Midpoint of the global profile (Fig.~\ref{fig2}) at 50\% of the peak.\hfill\break
$b)$ Velocity width at 50\% of the global profile peak (Fig.~\ref{fig2}).\hfill\break
$c)$ Continuum flux densities.  Upper limits (3$\sigma$) assume an unresolved point source.\hfill\break
$d)$ For NGC~5433, quantities derived from total channel fluxes have both random and systematic errors (denoted {\it r} and {\it s}, respectively), the latter a result
of RFI excision. See \S\ref{obsred} and Appendix~\ref{rfi} for details.\hfill\break
\end{minipage}
\end{table*}

In Table~\ref{global_params}, we present global properties of NGC~5433 obtained from
the C array maps.  The systemic velocity $V_{sys}$, measured line width at 50 per cent of the peak $W_o\sin{i}$, and integrated flux ${\int}S\,dV$ are in rough agreement 
with previously published values from single-dish measurements
(Mirabel \& Sanders 1988, Schneider et al. 1990).  This suggests that we have recovered
 all of the \ion{H}{1} emission and we thus make no 
short-spacing corrections to ${\int}S\,dV$ or in our galaxy models (see \S\ref{model}). The \ion{H}{1} mass is computed assuming an optically thin disc as in Shostak (1978), where $D$ is the distance to NGC~5433 in Mpc:
\begin{equation}
\label{MHI}
M_{HI}=2.36\,{\rm{x}}\, 10^{5}D^2 {\int}S\,dV \,\,\,\rmn{M_{\odot}}.
\end{equation}
The total dynamical mass is computed assuming a spherical dark matter distribution:
\begin{equation}
\label{MT}
M_{T}=(6.78\mbox{ x }10^{4})\theta_r\,D\,V^{2}(r)\,\,\,\rmn{M_{\odot}}. 
\end{equation}
In Equation (\ref{MT}), $\theta_r$ is the disc radius in arcmin and $V(r)$ is the
inclination-corrected rotation velocity. We make no correction for turbulence, as it is expected to be much smaller than the quoted 
uncertainties (Tully \& Fouqu\'e 1985).
For NGC~5433, we estimate $M_T$ for the approaching and receding sides of the galaxy separately adopting as $\theta_r$ and $V(r)$ the lowest contour in
Fig.~\ref{fig5}, and average the result to obtain the value in Table~\ref{global_params}.

\begin{figure*}
\includegraphics{fig4.ps}
\caption
{
(a) Total intensity map of NGC~5433 (contours; from C array
uniformly weighted data) overlaid on an optical image (greyscale; DSS).  Contours 
are at 63.5~$\,\times\,$~(2, 3, 4, 5, 6, 7, 8, 9, 10) $\rm{Jy}\,\rm{m}\,\rm{s}^{-1}\rm{beam}^{-1}$.  These values would 
be multiplied by 6.70~$\times$~10$^{18}$ to convert to column density,
N$_{\rm HI}$,
in units of cm$^{-2}$.
(b) First moment map of NGC~5433, from the C data. Contours are in 20 $\rm{km}\,\rm{s}^{-1}$ increments.
\label{fig4}
}
\end{figure*}

In light of the recent morphological re-classification of NGC~5433 
to SAB(s)c: by Dale et al. (2000) 
from the original RC3 classification of Sdm, 
it is interesting to compare its global properties 
with those of a sample of normal field spirals of various morphological 
types. For this exercise we use the volume-limited RC3-LSc sample of Roberts \& Haynes (1994) as the comparison sample, and compute the physical parameters
of NGC~5433 using the single-dish data of Schneider et al. (1990) for
the \ion{H}{1} parameters (the results are the same if our synthesis data
are used).  The results in Table~\ref{RH94_1} show that,
although there is overlap in properties between the morphological 
types, the physical parameters of
NGC~5433 are indeed more typical of the newer Sc classification
 than for Sd or Sm.

\begin{table*}
 \centering
 \begin{minipage}{135mm}
  \caption{NGC~5433 and Typical Properties of Late-Type Spirals\label{RH94_1}}
  \begin{tabular}{@{}lcccccc@{}}
\hline
Galaxy & log($R_{lin}$)$^b$ & log(${L_B}$) & log(${M_T}$)
& log($\sigma_{HI}$)$^c$ & log($M_{HI}$)
& log($M_{HI}$/$L_B$)\\
(1) & (2) & (3)
& (4) & (5) & (6) & (7) \\
\hline
NGC~5433$^a$ & 1.18 $\pm$ 0.05 & 10.30 & 11.17 $\pm$ 0.05 & 0.84 $\pm$ 0.09 & 9.69 $\pm$ 0.05 & -0.61 $\pm$ 0.05 \\
Sc$^d$ & 1.13$\,\to\,$1.40 & 10.2$\,\to\,$10.9  & 10.8$\,\to\,$11.3 & 0.80$\,\to\,$1.27 & 9.8$\,\to\,$10.4 & -0.3$\,\to\,$-0.7 \\
Sd$^d$ & 0.85$\,\to\,$1.22 & 9.8$\,\to\,$10.3  & 10.2$\,\to\,$10.7 & 0.83$\,\to\,$1.22 & 9.4$\,\to\,$10.0 & 0.0$\,\to\,$-0.5 \\
Sm$^d$ & 0.88$\,\to\,$1.16 & 9.4$\,\to\,$10.2  & 9.9$\,\to\,$10.4 & 0.82$\,\to\,$1.28 & 9.3$\,\to\,$9.8 & 0.1$\,\to\,$-0.3   \\
\hline
\end{tabular}
$a)$ Parameters as in Roberts \& Haynes (1994), using single-dish data from Schneider et al. (1990).\hfill\break
$b)$ Linear radius (kpc) from RC3.\hfill\break
$c)$ \ion{H}{1} surface density in $\rmn{M_{\odot}}\,\rmn{pc}^{-2}$, defined as $\sigma_{HI}=M_{HI}/\pi R_{lin}^2$.\hfill\break
$d)$ Range of values (25th to 75th percentile) from RC3-LSc sample of Roberts \& Haynes (1994), adjusted to $H_o=70\,\rm{km\,s^{-1}\,Mpc^{-1}}$.\hfill\break
\end{minipage}
\end{table*}

\begin{figure}
\includegraphics{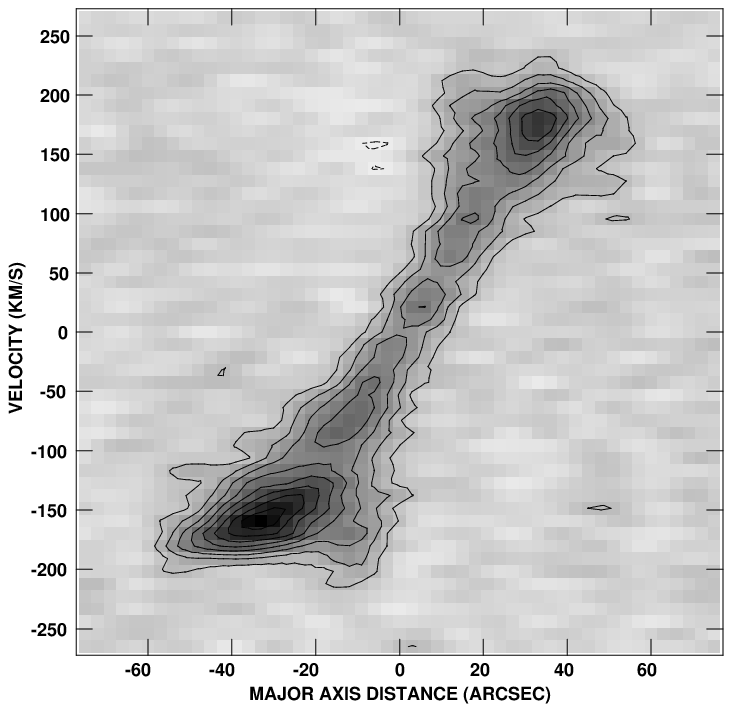}
\caption{3 arcsec--wide position-velocity slice of the uniformly weighted data shown in Fig.~\ref{fig4}, along the major axis of NGC~5433. Contours are at (-1, 1, 2, 3, 4, 5, 6, 7, 8, 9) $\rm{mJy}\,\rm{beam}^{-1}$, and greyscale runs from 0 to 9.08 $\rm{mJy}\,\rm{beam}^{-1}$. Coordinates are relative to optical centre and recessional velocity of NGC~5433 (Table~\ref{basic}).
\label{fig5}
}
\end{figure}

\subsubsection{Kinematic modelling}
\label{model}

Our \ion{H}{1} data for NGC~5433 contain intensity
measurements at 
many positions and velocities above the noise,
all of which provide important information about
the galaxy's global \ion{H}{1} distribution and dynamics.
Thus, rather than modelling the
moment maps alone, we have modelled the \ion{H}{1} spectra
at all positions in the cube.
  This is done by parameterizing the
velocity and density distributions in the galaxy,
specifying a geometry, producing 
 model \ion{H}{1} spectra, and then searching through
parameter space for the best fit.  
The benefits of this method are that all data in the cube
are used, the volume density distributions (rather than just
the column densities) and the system geometry can be constrained, and it is possible to disentangle the
effect of a thick disc from the effect of inclination on
the projected distribution perpendicular to the major axis.  
In addition, examination of the residual
cube 
can provide more insight into the peculiarities of
the \ion{H}{1} distribution in the galaxy.

The routine used (called {\sevensize CUBIT} and freely available from
{\tt http://www.astro.queensu.ca/$\,\tilde{\,}$irwin})
adopts a Brandt rotation curve and can
also apply a velocity dispersion if it is warranted.
The density distribution in the disc can be
either constant, decreasing exponentially, gaussian, or 
a gaussian ring with a centre at a specified radius
and different scale lengths on either side of this position. The latter distribution is useful in modelling galaxies with a dearth of 
\ion{H}{1} at small galactocentric radii.
The most recent version of the routine also allows for two
radial distributions to be superimposed.  
The vertical density distribution can be either
constant, exponential, or gaussian.
Details of this process can be found in Irwin \& Chaves (2003) and references therein, or at {\tt http://www.astro.queensu.ca/$\,\tilde{\,}$irwin}.  

For NGC~5433, we modelled the C configuration data at three different weightings/resolutions (natural, robust and uniform; see Table~\ref{observe}) to ensure that the results are independent of the imaging parameters.  This covers a range of spatial resolutions
from 13 -- 19 arcsec (4.1 -- 5.7 kpc).
The quoted  errors on each best-fitting parameter in the model
reflect the variation in results over the variously weighted cubes. 
Although there are up to 15 free parameters in the model, 
there are 
over 250 independent data points above 3$\sigma$ in the robustly weighted cube; thus the parameters are well constrained by the data.

\begin{figure}
\includegraphics{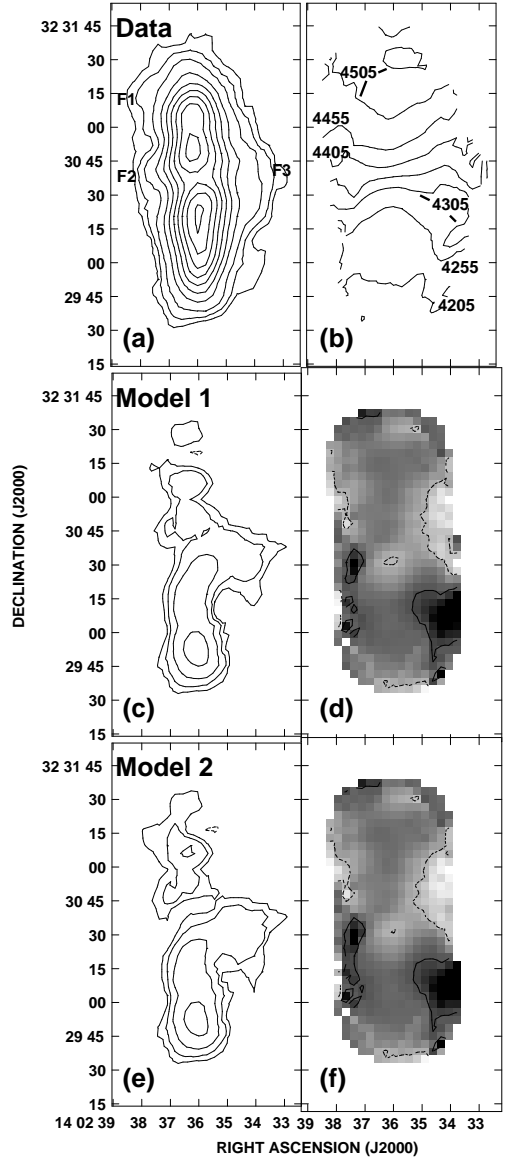}
\caption
{Moment maps of robustly weighted NGC~5433 data and residuals of corresponding models.  
 Total intensity contours may be converted to column densities N$_{\rm HI}$ (in units of cm$^{-2}$) by multiplying levels by 4.5~$\times$~10$^{18}$.
(a)  Integrated
intensity (zeroth moment map) of the data.  
 Contours are at 10~$\,\times\,$~(2.5, 5, 10, 20, 30, 40, 50, 60, 70, 80) $\rm{Jy}\,\rm{m}\,\rm{s}^{-1}\rm{beam}^{-1}$. 
 The features F1 -- F3 identified in Fig.~\ref{fig4} are labelled.
(b)  First moment map of the data.  
(c) Zeroth moment map made from the residual (data - model) 
cube for Model 1. Contours are at 
10~$\,\times\,$~(2.5, 5, 10, 20) $\rm{Jy}\,\rm{m}\,\rm{s}^{-1}\rm{beam}^{-1}$. 
(d) First moment map of the data in 6b
minus the first moment map made from Model 1.  The greyscale ranges
from  -50 (light) to +50 (dark) km s$^{-1}$ and contours are at -20 km s$^{-1}$ (dashed)
and +20 km s$^{-1}$ (dotted).  
(e) Same as in 6c, but for Model 2. 
(f) Same as in 6d, but for Model 2. 
\label{fig6}
}
\end{figure}

The best-fitting model values for NGC~5433 are given in 
Table~\ref{model_params} and moment maps made from the residuals 
(data minus model) are displayed in Fig.~\ref{fig6} 
for the robustly 
weighted cube. Model 1 gives the best
results for the case in which a single radial density distribution
is used, and Model 2 is the best fit for the sum of two radial density distributions.
For both models the RMS noise of the residual cube is less than 
3$\sigma$ of the input data, suggesting that the global 
morphology and kinematics of NGC~5433 are well--represented by the models.
 The forms of these distributions are given in the notes 
to Table~\ref{model_params}.

The first 5 parameters of Table~\ref{model_params} specify the  kinematic centre of the galaxy and the disc
geometry.  $V_{sys}$ agrees with the value from the global profile (Fig.~\ref{fig2}) but the kinematic centre is 
3.3 arcsec south of the optical centre, likely due
to the observed \ion{H}{1} asymmetry with more gas weighted
toward the south (Fig.~\ref{fig4}). 
The remaining parameters  
describe the 
shape of the rotation curve and the
density distributions within
and perpendicular to the midplane. 
In the midplane, the best-fitting distribution is 
a ring centred at $R_0=12.6$~kpc.
The RMS of the residual cubes, in the last row of Table~\ref{model_params},
shows that the addition
of the second radial density distribution peaking at
the centre with a characteristic radius of $D_c=3.5$ kpc
(Model 2) improves the fit only marginally. 
Our models therefore do not provide conclusive evidence for a central \ion{H}{1} depression in NGC~5433. 
For both Model 1 and Model 2, the vertical exponential scale
height ($H_e$ $\le$ 1.2~kpc) is an upper limit because the result is comparable
to the pixel size adopted for the cube.

The moment maps of the residual emission (Fig.~\ref{fig6})
clearly show the HI asymmetry mentioned previously, with
excess emission on the south side of the galaxy.
The remaining features in the residual emission
appear to be related to the extensions, discussed further in
the next section.
Note that these residuals can not be eliminated by varying the
model parameters, such as the input position angle (PA).

\begin{table}
\caption{Best-Fitting Kinematic and Density Parameters\label{model_params}}
\begin{tabular}{@{}lcc@{}}
\hline
Parameter & Model 1  & Model 2 \\
          & (1 Radial Dist.) & (2 Radial Dist.)  \\
 (1) & (2) & (3) \\
\hline
$\alpha$$^a$ (J2000) (h m s) & 14 02 36.00 $\pm$ 0.05 & 14 02 35.99 $\pm$ 0.06 \\
$\delta$$^a$ (J2000) ($^\circ$ $^\prime$ $^{\prime\prime}$)   
& 32 30 34.5 $\pm$ 0.5 & 32 30 34.5 $\pm$ 0.5 \\
PA$^b$ ($^\circ$) & 3.5 $\pm$ 0.4 &  3.1 $\pm$ 0.4 \\
i ($^\circ$) & 81.0 $\pm$ 0.2 &  83 $\pm$ 2\\
V$_{sys}$ (km s$^{-1}$) & 4356 $\pm$ 3 & 4355 $\pm$ 3  \\
V$_{max}$$^c$ (km s$^{-1}$) & 198.8 $\pm$ 0.2 & 198 $\pm$ 1\\
R$_{max}$$^c$ (kpc)                 & 6 $\pm$ 1 & 5.3 $\pm$ 0.3\\
m$^c$ & 2.0 $\pm$ 0.6 & 1.5 $\pm$ 0.4 \\
n$_0$$^d$ (cm$^{-3}$)& 0.139 $\pm$ 0.007 & 0.13 $\pm$ 0.01\\
R$_0$$^d$ (kpc)            &  12.6 $\pm$ 0.3 & 12.6 $\pm$ 0.3\\
D$_o$$^d$ (kpc)             & 2.18 $\pm$ 0.09 & 2.24 $\pm$ 0.03 \\
D$_i$$^d$ (kpc)            & 10.1 $\pm$ 0.6 & 10.1 $\pm$ 0.6 \\
D$_c$$^d$ (kpc)             & & 3.5 $\pm$ 0.3 \\
n$_c$$^d$  (cm$^{-3}$)        &  & 0.11 $\pm$ 0.01\\
H$_e$$^e$ (kpc)             & $\le$ 1.2 $\pm$ 0.1 &  $\le$ 1.2 $\pm$ 0.3\\
RMS$^f$ (mJy beam$^{-1}$)   & 0.49 &  0.45   \\ 
\hline
\end{tabular}
$a)$ Kinematic centre of the \ion{H}{1} distribution.\hfill\break
$b)$ PA of the receding major axis.\hfill\break
$c)$  Rotation curve parameters, defined by the Brandt curve:\hfill\break
$V(R)\,=\,V_{max}\,(R/R_{max}){[1/3\,+\,2/3\,{(R/R_{max})}^m]}^{(-3/2m)}$\hfill\break
\noindent where $R$ is the radial distance in the plane of the galaxy and $V(R)$ is the rotational velocity at that radius.\hfill\break
$d)$ The radial volume density distribution parameters are given by $n(R)\,=\,n_1(R)\,+\,n_2(R)$, where\hfill\break
$ n_1(R)\,=\,n_0\,exp[-(R-R_0)^2/{D_o}^2)]~~~~~ {\rm for}\,\,R>R_0,$\hfill\break
$ n_1(R)\,=\,n_0\,exp[-(R_0-R)^2/{D_i}^2)]~~~~~ {\rm for}\,\,R\le R_0,$\hfill\break
$ n_2(R)\,=\, n_c\,exp[-(R^2/{D_c}^2)].$\hfill\break
$e)$ Perpendicular exponential scale height.  The density at any point in the galaxy is $n(R,z)\,=\,n(R)\,exp(-z/H_e)$, where $z$ is the perpendicular distance from the plane of the galaxy.\hfill\break
$f)$ RMS of the residual (model-data) cube after applying a 2$\sigma$ cutoff, where $\sigma$ is the RMS noise level of the original input data.\hfill\break
\end{table}

\subsubsection{High-latitude \ion{H}{1} in NGC~5433}
\label{extensions}


The residual maps in Fig.~\ref{fig6} reveal departures from the 
best-fitting models of NGC~5433 at high latitudes, coincident with the
features F1 -- F3 in the total intensity maps 
(Figs.~\ref{fig4} and \ref{fig6}a).  We examine these extensions in the C configuration data, which is most suited to the
analysis given its spatial and spectral resolution.
 Since NGC~5433 is not truly edge-on ($i=78$\degr, Table~\ref{basic}) and since many features of interest are comparable in size to the map resolution, the linear dimensions we infer from projected angular sizes in the map are only approximations to the true physical scales involved. 

\begin{figure*}
\includegraphics{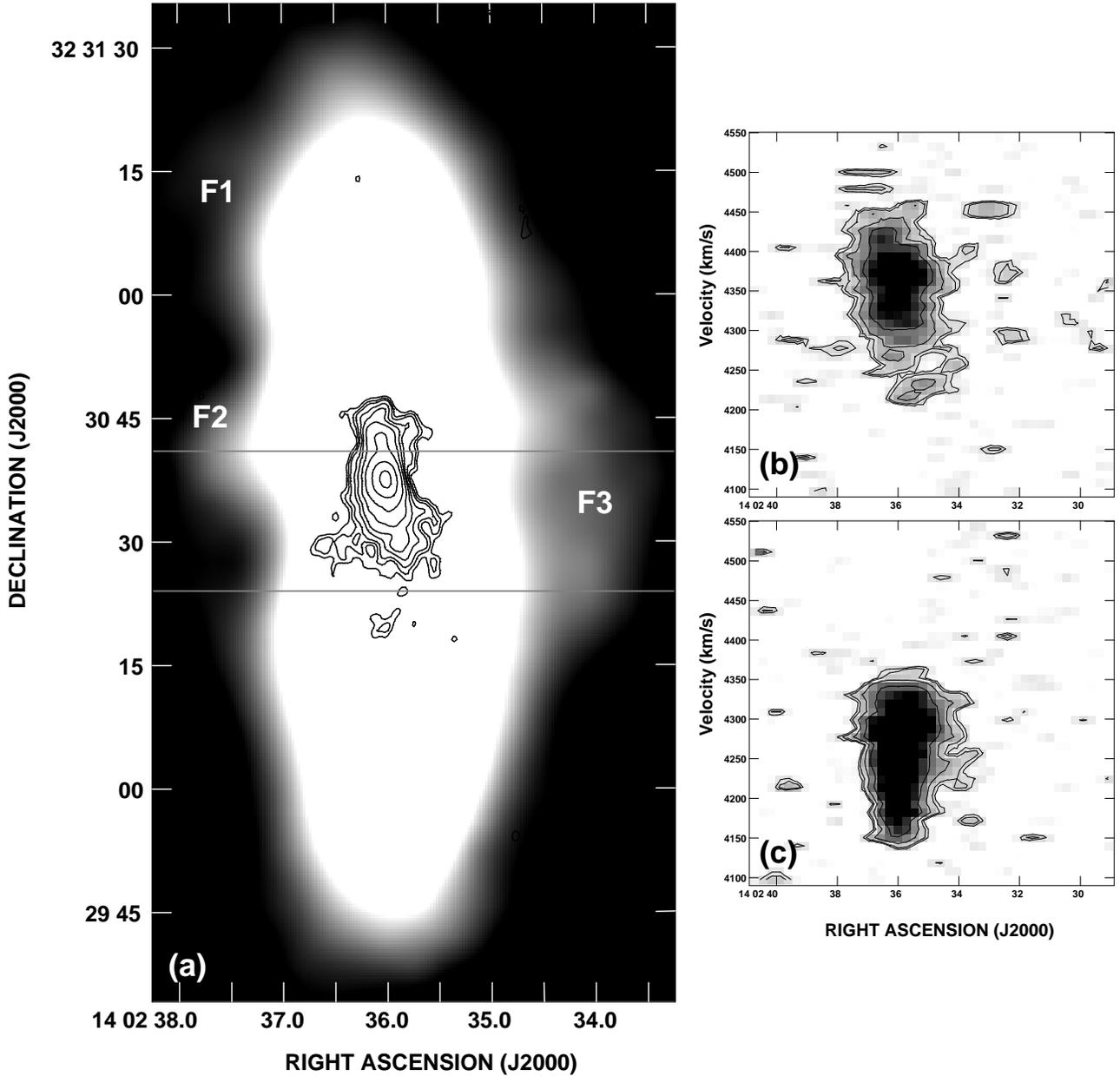}
\caption
{(a) High resolution continuum map of NGC~5433 (contours) taken from
Irwin et al. (2000), superimposed on an HI zeroth moment map (greyscale,
same as Fig. 6a) with the greyscale range from 50 to 250 Jy m s$^{-1}$
beam$^{-1}$. Horizontal lines show locations of P-V slices in 7b and 7c. (b)  P-V cut taken along the top horizontal
line in 7a. Contours are at 0.6, 0.75, 1.2, 2.0, and 4.0 
mJy beam$^{-1}$ and the greyscale is from 0.3 to 3.9 mJy beam$^{-1}$.
 (c)  Same as 7b, but for a P-V cut along the
bottom horizontal line in 7a.
\label{fig7}
}
\end{figure*}

In Fig.~\ref{fig7}a, we 
show a total intensity \ion{H}{1} map of the robustly weighted datacube for NGC~5433, 
enhanced to illustrate low intensity extensions, with the
high resolution radio continuum emission contours of Irwin et al. (2000) superimposed. 
Extensions from the mid-plane are clearly present in both ISM tracers.  
As has been observed in other galaxies (e.g. Collins et al. 2000), when extensions exist, they can be seen in a variety of components.
The low intensity emission shows that the 
\ion{H}{1} features extend as far as 25 arcsec  ($\sim$5 kpc) 
in projection from the midplane.  This is greater than the disc thickness 
($\le$ 1.2 kpc, \S\ref{model}) but appears to be typical of galaxies
showing \ion{H}{1} extensions (e.g. Lee \& Irwin 1997).
Feature F3, which appears as a broad
single feature in Fig.~\ref{fig6}a, is revealed to consist of 
at least two components, one to the north and one to the south of the nucleus. The continuum image shows two similar features on the west side, despite its drastically different resolution of 1.5 arcsec compared to that of the \ion{H}{1} data.  Given the correspondence of the
western extensions in the maps, it may be that each continuum
 feature is related to the nearest \ion{H}{1} extension and possibly to each other as well.

It is not always clear whether HI extensions identified in total intensity maps represent coherent structures, or whether they consist
of unrelated features along the same line of sight.  To distinguish between these possibilities, we have searched through the data both in position-position (P-P) space as well as in position-velocity (P-V) space to look for coherent structures related to the features F1, F2, and F3.

In the case of F1 ($14^{\rmn{h}}\,02^{\rmn{m}}\,38^{\rmn{s}}$, $32^{\circ}\,31{\arcmin}\,15{\arcsec}$),
2 -- 5$\sigma$ 
extensions from the underlying disc are seen at 
4522 and 4490 -- 4469~$\rm{km\,s^{-1}}$ 
(Fig.~\ref{fig3a}). The features from 4490 -- 4469~$\rm{km\,s^{-1}}$ are also evident 
in the C+D data, and given the direction of rotation of NGC~5433, their NE orientation
 is consistent with the slight redshift of the data relative to the 
best-fitting disc models in this vicinity (Figs.~\ref{fig6}d and~\ref{fig6}f).
  If they represent a single, coherent structure, then using the fluxes measured for these extensions over the 4490 -- 4469~$\rm{km\,s^{-1}}$  velocity range in Equation~(\ref{MHI}) we compute $M_{HI}\sim7\times10^7\,M_{\odot}$. This is broadly
consistent with $M_{HI}$ estimated for F1 from the total intensity distribution in Fig.~\ref{fig4}a. 

At the position of F2 ($14^{\rmn{h}}\,02^{\rmn{m}}\,37\fs 5$, $32^{\circ}\,30{\arcmin}\,40{\arcsec}$) there are many features
 in the channel maps, such as 2 -- 3$\sigma$ peaks from 
4447 -- 4426~$\rm{km\,s^{-1}}$ 
and a series of extensions from 4331 -- 4299~$\rm{km\,s^{-1}}$.
However, there is no evidence that these features
form a coherent structure.  This is illustrated in the P-V slice
shown in Fig.~\ref{fig7}b: near F2 (corresponding 
to a vertical strip to the left of the main emission), there are a number of small, likely unrelated extensions 
at a variety of velocities which collectively contribute to F2. 

F3 is located at $14^{\rmn{h}}\,02^{\rmn{m}}\,34^{\rmn{s}}$, $32^{\circ}\,30{\arcmin}\,35{\arcsec}$ 
(approximate midpoint).
We see 2 -- 3$\sigma$ extensions in 
Fig.~\ref{fig3a} at that location at 4437, 4405, 4394 and 
4352~$\rm{km\,s^{-1}}$, and 2 -- 5$\sigma$ ones 
from 4256 -- 4235~$\rm{km\,s^{-1}}$. The latter 
are also present in the C+D data, and are consistent with the residual maps in Figs~\ref{fig6}d and~\ref{fig6}f.  
Considering features in this velocity range to
represent a coherent structure,
we compute $M_{HI}\sim 1.1\times10^8\,\rm{M_\odot}$ from Equation (1), 
 again consistent with the mass computed for F3 in Fig.~\ref{fig4}a.

The P-V slice through
the top part of F3 (Fig.~\ref{fig7}b, right side of
main emission) further illustrates that the
emission does appear to form
a coherent structure which is roughly circular, with the centre at $\alpha$$\,\sim14^{\rmn{h}}02^{\rmn{m}}35^{\rmn{s}}$, $V\sim4250\,\rm{km\,s^{-1}}$:
this is a signature of an expanding shell or part thereof.  
The diameter of the shell-like feature is 50 km s$^{-1}$ in velocity
and 25 arcsec ($\sim$3 kpc) in right ascension.
An estimate of the \ion{H}{1} mass from this P-V slice only
is $\sim3\times10^7\,\rm{M_\odot}$. Under the assumption that the shell is expanding, 
we thus compute  an expansion
kinetic energy $\sim 2\times10^{53}$ ergs,
clearly a lower limit for that associated with the feature as not all the mass is represented
in this slice.
A P-V slice through the lower part of F3 (Fig.~\ref{fig7}c) shows more emission at a
similar velocity, and is typical of what is seen in other
slices through this feature.  If all of the mass computed for this
coherent part of F3 ($1.1\times10^8\,\rm{M_\odot}$) is involved
in the expansion, then the associated kinetic energy of the \ion{H}{1}
 is $\sim 8\times10^{53}$ ergs.

If the feature in Fig.~\ref{fig7}b is generated from a shperically symmetric, point injection of
energy such as a supernova, then the required input energy can be computed using the results of numerical simulations by Chevalier (1974) for an expanding shell that is now in the radiating phase of its evolution:
\begin{equation}
E\,=\,5.3\,\times\,10^{43}\,{n_1}^{1.12}\,{R_{sh}}^{3.12}\,{V_{sh}}^{1.4}~~~~{\rm ergs},
\label{injection}
\end{equation}
where $V_{sh}$ (km s$^{-1}$) and $R_{sh}$ (pc) are the expansion velocity and radius of the shell, and
$n_1$ (cm$^{-3}$) is the ambient density at the location of generation.
Using $V_{sh}$ and $R_{sh}$ of the structure in Fig~\ref{fig7}b and estimating $n_1$ at its location
from the models in \S\ref{model}, 
then from Equation~(\ref{injection}) $E\sim2\times10^{54}$ ergs.  
This estimate decreases only by a factor of 2 if, instead, a continuous deposition of energy is assumed (Weaver et al. 1977).

One alternative is to interpret the apparently 
circular structure of Fig.~\ref{fig7}b as disconnected
components, pushed or pulled out of the midplane and collectively contributing to F3.  
However, even if the clouds aren't related one still
needs to place $\sum M_{cloud} \sim10^8\,\rm{M_\odot}$ an average distance $z\sim2.5$~kpc
above the plane.  If $z_0$ is
the vertical scale height of the stellar mass distribution and $\rho_0$ is the stellar mass density at the midplane,
then the potential energy of a ``cloud'' in equilibrium at a distance $z$ above a thin, axisymmetric disc is
\begin{equation}
\Phi(z)=2\times10^{54}\,\big(\frac{M_{cloud}}{2\times10^7\rmn{M}_\odot}\big)
\,\big(\frac{z_0}{700\,\rmn{pc}}\big)^2
\,\big(\frac{\rho_0}{0.185\,\rmn{M_{\odot}pc^{-3}}}\big)
\,\ln{\big[\cosh \big( \frac{z}{z_0} \big) \big]}\,\,\,\,\rmn{ergs}
\label{potential}
\end{equation}
(c.f. Binney \& Tremaine 1987, Lee et al. 2001), where we normalize $z_0$ to Galactic and $\rho_0$ to solar neighbourhood values, respectively. Adopting these values for $z_0$ and $\rho_0$ we find $\Phi\sim3\times10^{55}$ ergs for $\sum M_{cloud}=10^8\,\rmn{M}_\odot$ and $z=2.5$~kpc, 
 larger than the injection energy computed above.
While these energy estimates are order-of-magnitude approximations (c.f. Lee \& Irwin 1997), 
it is clear that the energy required to produce F3 is very high, and comparable to inferred values for the largest supershells in 
the Galaxy (McClure-Griffiths et al. 2002) and those in other edge-on spirals (e.g. Chaves \& Irwin 2001).

\subsection{The environment of NGC~5433}
\label{HI_environment}

To characterize the environment of NGC~5433, we imaged the C+D data with a variety of weighting schemes to emphasize structures on different spatial scales, and systematically searched the resulting \ion{H}{1} cubes. In all, there are 20 galaxies within a 16 arcmin radius (the primary beam HWHM) of NGC~5433 with MAPS or 2MASS identifiers. Of these, we detect \ion{H}{1} in 3 galaxies other than NGC~5433 itself: KUG~1359+326, SIS-1 and SIS-2 (see Fig.~\ref{fig1}). We note again here that IC~4357, a confirmed companion 44 arcmin SW of NGC~5433 (Giuricin et al. 2000), is not within the primary beam of our observations; we do not search for emission outside this region. We find no other candidates with peak fluxes over 4$\sigma$ spanning more than 2 consecutive channels. Adopting these parameters as detection criteria, our maps are sensitive to $M_{HI} = 4 \times 10^7\, M_{\odot}$ at the distance of NGC~5433.

\begin{figure*}
\includegraphics{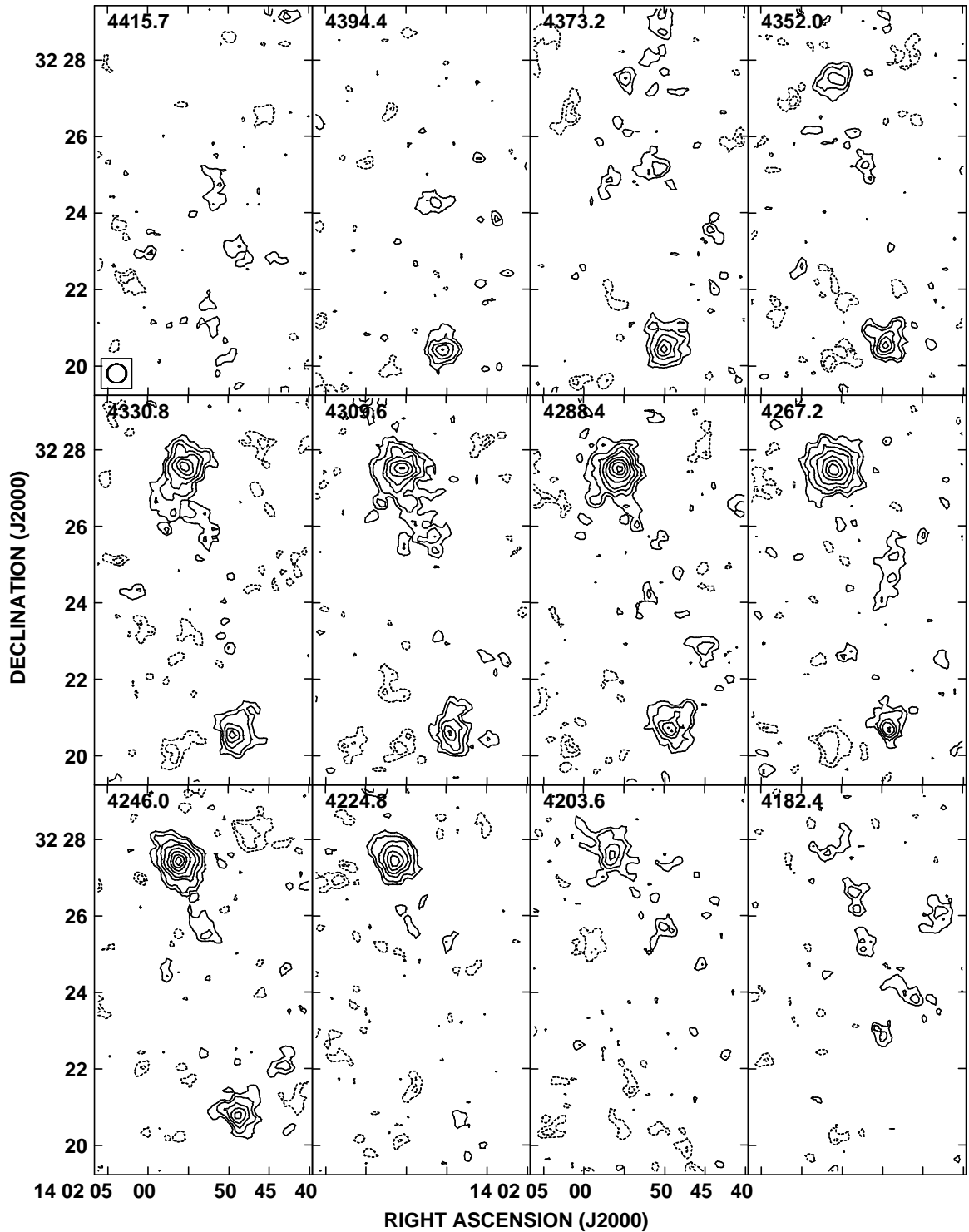}
\caption
{Naturally weighted C+D channel maps for KUG~1359+326 (in the 
upper half of the frames) and SIS-1 (in the lower half).  Contours are at 
0.21~$\,\times\,$~(-3 (dashed), -2 (dashed), 2 (2$\sigma$), 3, 4.5, 6, 8, 10, 12, 14, 16) $\rm{mJy}\,\rm{beam}^{-1}$. The 
channel velocity in $\rmn{km}\,\rmn{s}^{-1}$ is indicated in the upper right corner of each frame, and the synthesized beam is given in the lower left corner of the first frame.
\label{fig8}
}
\end{figure*}

 The detection of \ion{H}{1} in KUG~1359+326
 (projected separation = 200 kpc), confirms 
that this galaxy is indeed a physical
companion to NGC~5433.  We do not detect CGCG~191-037, despite its proximity to NGC~5433 (projected 
separation = 102 kpc), diameter intermediate to NGC~5433 and the companion KUG~1359+326, and blue magnitude comparable to the latter (Table 1). Its near-infrared brightness is also intermediate to that of these two group members, as it is an average of $\sim$1.25 magnitudes fainter than NGC~5433 and $\sim$1.27 magnitudes brighter than KUG~1359+326 across the 2MASS J, H and K bands.  Perhaps this galaxy's recessional velocity places it outside our 657 $\textrm{km}\, \textrm{s}^{-1}$ bandpass centred on NGC~5433. In this scenario CGCG~191-037 is at least 4.5 Mpc from NGC~5433, if their relative velocity is dominated by Hubble flow. While the size and magnitudes of CGCG~191-037 are comparable to those of KUG~1359+326, its magnitude relative to the dominant group member NGC~5433 suggest that it is a background (rather than a foreground) object.
Alternately, CGCG~191-037 may have a substantial peculiar velocity relative to NGC~5433, or may be devoid of \ion{H}{1} altogether. We also searched for continuum emission in CGCG~191-037, and found none. If our non-detection of CGCG~191-037 in \ion{H}{1} stems from this galaxy being closer or farther from NGC~5433 than its optical dimensions suggest, the continuum non-detection supports the latter possibility. 

Particularly interesting is the detection of two new companions, SIS-1 and SIS-2.
SIS-1  is a 17.5th magnitude galaxy with a discernible disc, and its projected
 separation from NGC~5433 is large at 289 kpc.
SIS-2 is an 18th magnitude galaxy just south of NGC~5433, separated from its centre
by 2.9 arcmin (55 kpc). It is clear that NGC~5433 is the dominant member of a group of at least 4 galaxies (5 if IC~4357 is included).  At the sensitivity of our observations, all the \ion{H}{1} companions within 300 kpc of NGC~5433 are to the SW of the latter.  


\begin{figure}
\includegraphics{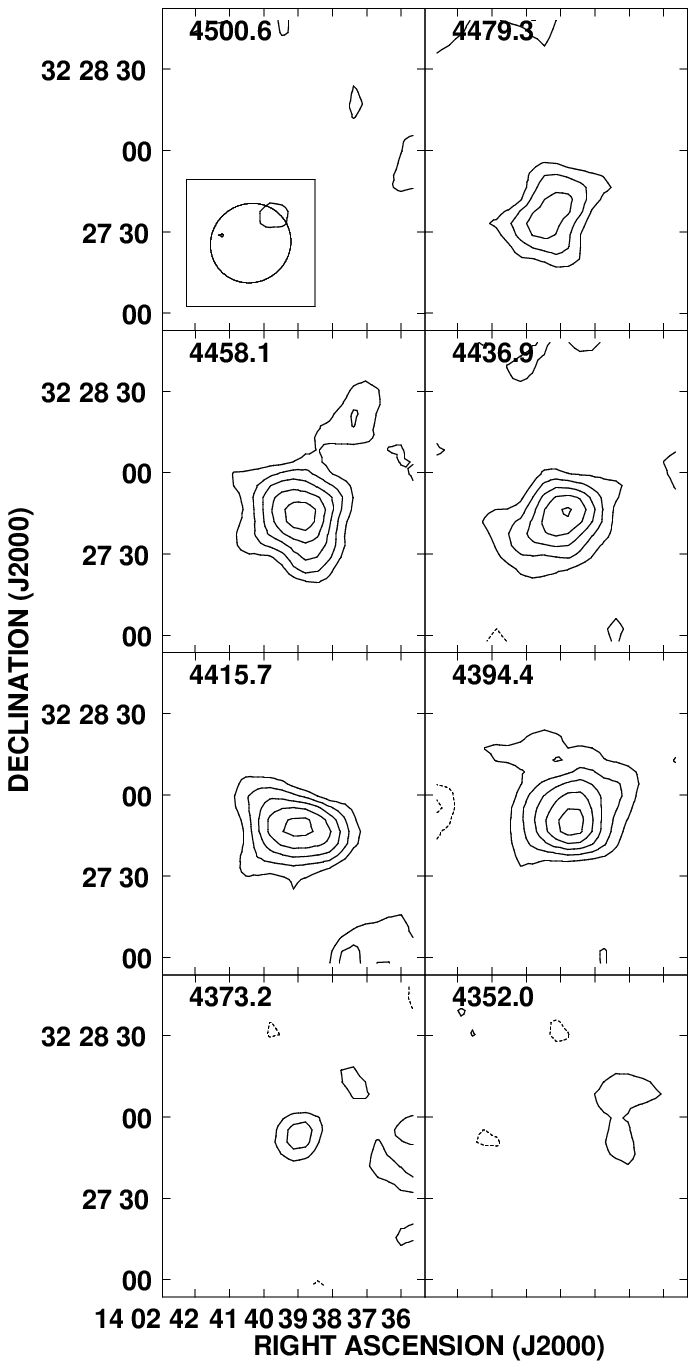}
\caption
{Naturally weighted C+D channel maps for SIS-2.  Contours are at 
0.21~$\,\times\,$~(-3 (dashed), -2 (dashed), 2 (2$\sigma$), 3, 4, 5, 6, 7) $\rm{mJy}\,\rm{beam}^{-1}$. The 
channel velocity in $\rmn{km}\,\rmn{s}^{-1}$ is indicated in the upper right corner of each frame, and the synthesized beam is given in the lower left corner of the first frame.
\label{fig9}
}
\end{figure}

\subsection{\ion{H}{1} in the companions}
\label{HI_companions}

Although the companions of NGC~5433 are detected in both our C and C+D maps, we analyse of their \ion{H}{1} distributions using the higher sensitivity C+D data. 
Naturally weighted C+D channel maps for KUG~1359+326, SIS-1 and SIS-2 are shown in Figs.~\ref{fig8} and \ref{fig9}, and the global profiles derived from these maps are found in Fig.~\ref{fig2}. Moment maps for each galaxy, again obtained from the naturally weighted C+D data, are given in Fig.~\ref{fig10}.
The global \ion{H}{1} properties are listed in Table~\ref{global_params}. In computing $M_{HI}$ and $M_T$ from Equations~(\ref{MHI})~and~(\ref{MT}), we assume that the companions are at the same distance as NGC~5433. For $M_T$, we approximate $\theta_r = 1.5 \times a/2$ where $a$ is the optical major axis diameter of each system (Table~\ref{basic}).
Given the resolution of these observations and kinematic distortions observed (see below), 
$M_T$ only provides an indication of the true dynamical masses of the companions.
 
 We discuss the \ion{H}{1} properties of each galaxy below. Because of the low spatial and spectral resolution of the detections, we do not attempt to model their \ion{H}{1} distributions.

\subsubsection{KUG~1359+326}
\label{KUG}

Our strongest detection among the companions, KUG~1359+326 shows \ion{H}{1} emission from 4373 -- 4204~$\rm{km\,s^{-1}}$ (Fig.~\ref{fig8}), and is resolved in most of these channels. The NW side of the galaxy is receding, and the SE side advancing. The most interesting feature of the maps is an extension apparently emanating from the SE part of the disc, primarily from 4331 -- 4288~$\rm{km\,s^{-1}}$. It extends $\sim2$ arcmin from the optical centre of the galaxy, and farther if the low-level emission at 4267 and 4182~$\rm{km\,s^{-1}}$ is also associated with the extension. Despite the low S/N of this feature, we detect it in both the C and D data independently: this strongly suggests that it is indeed real, rather than a chance thermal fluctuation in one of the datasets. The extension is also seen at a low level in the total intensity map in 
Fig.~\ref{fig10}a. At the distance NGC~5433 (and after correcting for beam smearing), emission is found over 30~kpc away from the disc of KUG~1359+326, over 4 times the optical major axis diameter of the galaxy. There is no coherent velocity field associated with the extension in our data, and as such it is not shown in Fig.~\ref{fig10}b. The origin of this extended feature is thus unclear from these data.

The first moment map of KUG~1359+326 (Fig.~\ref{fig10}b)
indicates that the velocity field of this galaxy is highly distorted. Most of the gas from 4270 -- 4300~$\rm{km\,s^{-1}}$ appears to lie along the optical major axis of the system, although the shape of the isovelocity contours indicates the presence of kinematic distortions along the minor axis at these velocities. The emission at the NE edge of the galaxy, coincident with a low-level extension in the same direction in Fig.~\ref{fig10}a, shows the most distorted velocity contours, which appear to be `pushed' northwards relative to those expected for an inclined spiral.


\subsubsection{SIS-1}
\label{SIS-1}

The channel maps for SIS-1 (Fig.~\ref{fig8}) show emission from 4394 -- 4246~km~s$^{-1}$, with the NW side of the galaxy advancing and the SE side receding. The low-level emission in the channel maps is barely resolved by our synthesized beam, and shows a number of distortions. Two of these features, extending to the NE and NW of the emission centroid primarily at 4352~km~s$^{-1}$, are also evident in the total intensity maps in Fig.~\ref{fig10}c. The NW feature appears as an extension from the tip of the major axis of the galaxy. The higher level contours in the zeroth moment map are slightly elongated along the optical major axis of SIS-1, their centroids roughly coincident with the optical peak. The velocity field in Fig.~\ref{fig10}d shows ordered rotation in the SE part of the disc, with the kinematic major and minor axes roughly aligned with the optical ones. The NW side appears to be kinematically distinct from the remainder of the disc, and associated with the NW feature of the total intensity map.

\subsubsection{SIS-2}
\label{SIS-2}

\begin{figure*}
\includegraphics{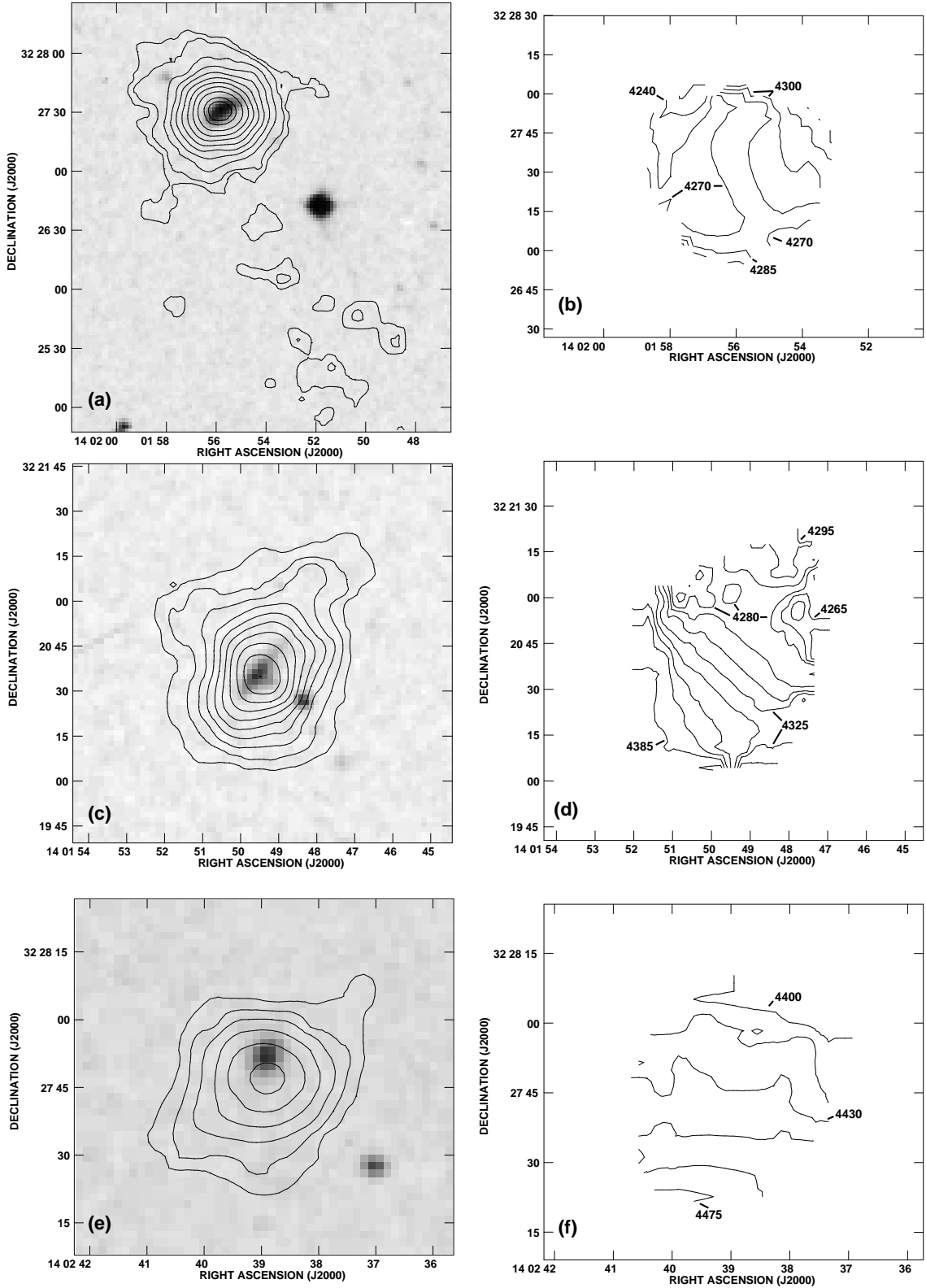}
\caption
{Total intensity \ion{H}{1} maps of detected galaxies 
superimposed on optical DSS images (left
hand panels) and their corresponding first moment maps (right hand panels).  Zeroth moment map contour values would 
be multiplied by 1.31~$\times$~10$^{18}$ to convert to column density,
N$_{\rm HI}$,
in units of cm$^{-2}$. All first moment map contours are separated by 15 km s$^{-1}$ unless otherwise labelled.
{\bf (a)}  Total intensity map of KUG~1359+326.  
Contours are at 39.4~$\,\times\,$~(0.5, 1, 2, 3, 4, 5, 6, 7, 8, 9, 9.5)
 $\rm{Jy}\,\rm{m}\,\rm{s}^{-1}\rm{beam}^{-1}$. 
{\bf (b)} First moment map of KUG~1359+326.
{\bf (c)} Total intensity map of SIS-1.  
Contours are at 24.9~$\,\times\,$~(1, 2, 3, 4, 5, 6, 7, 8, 9, 10) $\rm{Jy}\,\rm{m}\,
\rm{s}^{-1}\rm{beam}^{-1}$.
{\bf (d)} First moment map of SIS-1.  
{\bf (e)} Total intensity map of SIS-2.  
Contours are at 15.0~$\,\times\,$~(1.3, 3, 4.5, 6, 7.5, 9, 10) $\rm{Jy}\,\rm{m}\,
\rm{s}^{-1}\rm{beam}^{-1}$.  
{\bf (f)} First moment map of SIS-2.  
\label{fig10}
}
\end{figure*}

SIS-2 is the weakest source that we detect in the vicinity of NGC~5433, extending from 4479 -- 4373~km~s$^{-1}$ in Fig.~\ref{fig9}. It is none the less a definite detection in our data, with peak emission greater than 6$\sigma$ over four consecutive channels in the C+D maps as well as independent detections in the C and D datasets. The north side of the galaxy is advancing and the south side receding, and thus its rotation is retrograde relative to that of NGC~5433. We do not resolve the emission in most of the channels. As such, the total intensity map in Fig.~\ref{fig10}e is mostly featureless except for an offset in the \ion{H}{1} emission peak from the optical centre by about 7 arcsec, or $1/6$ of a beam. The velocity field of SIS-2 (Fig.~\ref{fig10}f) shows largely ordered, solid-body rotation throughout the disc, likely a result of the poorly resolved emission regions in each channel.

\section{Discussion}
\label{discussion}

 Our observations probe the \ion{H}{1} morphology and kinematics of NGC~5433 and its environment with unprecedented detail. In light of our results we discuss the prevalence of high-latitude \ion{H}{1} in edge-on spirals in \S\ref{disc1}, characterize the NGC~5433 system as an interacting group of galaxies in \S\ref{disc2}, and underscore the possible importance of environmental effects in driving the disc-halo phenomenon in \S\ref{disc3}.  


\subsection{High-latitude \ion{H}{1} in NGC~5433 and other spirals}
\label{disc1}
 We identify 3 extraplanar features in NGC~5433 by examining moments of the \ion{H}{1} distribution and best-fitting kinematic models. 2 of these (F1 and F3) appear to be associated with coherent features in P-P and P-V space (Figs.~\ref{fig3a}, \ref{fig7}); we find a complete P-V loop in the upper part of feature F3 (Fig.~\ref{fig7}b), which may represent an expanding shell. For a point injection of energy, the implied input energy for F3 is $E \sim 2 \times 10^{54}$ ergs (Equation~(\ref{injection})). This is nearly an order of magnitude larger than the corresponding value $E=4.6 \pm 2.0 \times 10^{53}$ ergs for the most energetic new supershell (GSH 292-01+55) found in the Southern Galactic Plane Survey (McClure-Griffiths et al. 2002), but lower than $E$ inferred for supershells found in other spirals, such as feature F1 in NGC~2613 ($E > 1.4 \times 10^{55}$ ergs, Chaves \& Irwin 2001) and F2 in NGC~5775 ($E \sim 2.1 \times 10^{55}$ ergs, Lee et al. 2001). 
Given the order-of-magnitude uncertainties on these estimates (Lee \& Irwin 1997), it is clear that the implied energetics for the feature F3 in NGC~5433 are comparable to those of the largest structures found in the Milky Way as well as features in other spirals. 

While the sample remains small, discrete \ion{H}{1} features and/or thick \ion{H}{1} discs have now been observed in galaxies with a wide range of star formation rates (e.g. higher: NGC~891, Swaters, Sancisi \& van der Hulst 1997; NGC~2403, Schaap et al. 2000; low: UGC~7321, Matthews \& Wood 2003)  and in a variety of environments (e.g. isolated: NGC~3044, Lee \& Irwin 1997; NGC~3556, King \& Irwin 1997; companions: NGC~5433; interacting: NGC~5775; Irwin 1994). Regardless of its origin, coherent extraplanar \ion{H}{1} emission may thus be common in spiral galaxies, as is the case for high-latitude cosmic-ray haloes (Irwin et al. 1999). The existence of these features implies significant energy transfers from the disc into the halo, by mechanisms that remain poorly understood. We return to plausible formation scenarios for features in NGC~5433 in \S\ref{disc3}.

\subsection{An active environment}
\label{disc2}
NGC~5433 is in a richer environment than
previously thought, having at least 3 physical companions (KUG~1359+326, SIS-1, SIS-2) rather than the 2
inferred through proximity and size alone (KUG~1359+326, CGCG~191-037).
We suspect that other faint MAPS galaxies
in the field might also be detectable in \ion{H}{1} with observations
of greater sensitivity.
Since we do not detect either \ion{H}{1} or 20~cm continuum emission from CGCG~191-037, its relationship to the detected galaxies remains unclear; \ion{H}{1} observations with a wider bandpass would resolve this issue.  
 If the more distant NOG companion IC~4357 is included, the group size increases to 5; \ion{H}{1} observations of IC~4357 of similar sensitivity to those presented here may also reveal physical companions, making the group richer still. 
We note that the X-ray detection reported for NGC~5433 (Rephaeli et al. 1995) may represent hot intergalactic gas, sometimes found in poor groups (Fukazawa et al. 2002).

It is interesting that all confirmed group members differ in velocity
from NGC~5433 by no more than 70 km s$^{-1}$ (Table~\ref{global_params}).
SIS-1, for example, at a
projected separation of 248 kpc, is separated in velocity from NGC~5433 by only 25~$\rm{km\,s^{-1}}$.  
Thus the projected separation of the galaxies may be close to their true separation. 
In addition, within 20 arcmin of NGC~5433 there are over 3 times as many 2MASS galaxies (12 vs. 4) and 2 times as many optically detected galaxies (14 vs. 7) in the NE+SW quadrants than in the NW+SE quadrants.
We thus speculate that the NE - SW orientation of the galaxies in this study may represent the largest systems in 
 a ``true" filament at least $770$~kpc in size (the separation between NGC~5433 and IC~4357), with little variation in depth. Thus the group geometry is in qualitative agreement with galaxy structures on larger scales found in surveys (e.g. Cross et al. 2001) and simulations of hierarchical structure formation via gravitational collapse (e.g. Bertschinger 1998).  


There is evidence that interactions are taking place in the group, either among the detected galaxies or with a hot intergalactic medium.  
NGC~5433, itself, shows some optical curvature (Fig.~\ref{fig1}) and has an
asymmetric \ion{H}{1} distribution
 (Fig.~\ref{fig2}), with more \ion{H}{1} near the close companion SIS-2.  
All of the detected companions show peculiarities in their \ion{H}{1} morphology. 
The channel maps of KUG~1359+326 reveal a large extension that could be tidal in origin (Figs.~\ref{fig8},~\ref{fig10}a), and the galaxy's velocity 
field is highly distorted (Fig.~\ref{fig10}b). Moment maps of SIS-1 (Figs.~\ref{fig10}c,~\ref{fig10}d) show a kinematically distinct feature roughly aligned with the galaxy's 
major axis, and the \ion{H}{1} peak in SIS-2 is offset from the optical centre (Fig.~\ref{fig10}e). Indeed, the better we resolve a companion to NGC~5433, the more distorted its \ion{H}{1} morphology and kinematics appear. This leads us to suspect that SIS-2 may indeed be strongly interacting with nearby NGC~5433, but that evidence of the encounter is concealed by our poor spatial and spectral resolution of this companion. 

\subsection{Underlying mechanisms}
\label{disc3}
Since NGC~5433 is a starburst (see fig. 2 of Irwin et al. 1999) it seems plausible that the observed high-latitude extensions are generated by an internal mechanism.
 A number of scenarios are consistent with the data. Symmetries in features about a galaxy midplane have previously been attributed to a single event in the underlying disc (Chaves \& Irwin 2001). If this is the case for F2 and F3 (Fig.~\ref{fig4}a), it is puzzling that we do not find extensions corresponding to F2 in the same channels as those harbouring the shell attributed to F3 (Fig.~\ref{fig7}b). 
On the other hand, the double peak in F3 itself and the presence of 20 cm continuum loops in the same vicinity (Fig.~\ref{fig7}a) suggest a common origin for all of these features; their locations relative to the galaxy nucleus give the impression of a broad-scale nuclear outflow on the west side of the galaxy. Alternatively, the apparent correlation between the continuum and \ion{H}{1} features is also consistent with a galactic fountain generated by energetic events in the disc (e.g. Lee et al. 2001). 


 None the less, one cannot dismiss the role of NGC~5433's active environment in the generation of the detected extraplanar features. For example, observations of 3 systems with low mass companions by van der Hulst \& Sancisi (2003) all show evidence for gas accretion on to the parent discs. Conversely, tidal interactions may pull gas out of the midplane of large spirals, sometimes forming arcs and plumes with similar morphologies to supershells (Taylor \& Wang 2003). In the Galaxy, the origins of some high-velocity clouds (HVCs) have been attributed to these processes (e.g. Putman et al. 2003). We speculate that a similar scenario applies to NGC~5433: given the inferred upper limits on the masses of HVC analogues in external systems (e.g. Pisano et al. 2004), they would not be detectable in our data. Tidal interactions between NGC~5433 and its smaller companions may be responsible for the observed extraplanar emission, either via direct impacts, tidally induced spurs, or by fuelling the increased star formation found in the disc. Additional multi-wavelength observations, particularly high-resolution X-ray data as well as high-sensitivity \ion{H}{1} and  H$\alpha$ maps, are needed to distinguish between these possibilities. 
Indeed, the environment can be a key component in the formation of {\it both} externally and internally generated high-latitude features.

\section{Summary}
\label{conclusions}

We have presented spatially resolved \ion{H}{1} observations of the edge-on starburst galaxy NGC~5433 and its environment. We find that NGC~5433 has an asymmetric \ion{H}{1} distribution with more gas in the southern half of the disc, and global properties typical of its new Sc classification. The main disc emission is best modelled with a gaussian ring at $R_o=12.6\pm0.3$ kpc in the plane with an exponential scale-height $H_e\le 1.2\pm 0.3$ kpc; the addition of a gaussian radial distribution with a characteristic width of $D_c=3.5\pm0.3$ kpc marginally improves the fit. 

 Further examination of the data and model residuals reveals 3 extraplanar features, that we label F1 -- F3. We identify coherent extensions in consecutive velocity channels corresponding F1 and F3, and find a complete loop in P-V space at the location of F3. Taking this as the signature of an expanding shell the required injection energy is $2 \times 10^{54}$ ergs, a value comparable to that found in galactic and other extragalactic supershells. Now detected in systems with a variety of star formation rates and in diverse environments, it appears that coherent extraplanar emission requiring such large input energies is a common phenomenon in spiral galaxies.


NGC~5433 is in a richer environment than previously thought. We confirm that KUG~1359+326 is indeed a physical companion to NGC~5433, and detect two other dwarf galaxies with MAPS identifiers in the field that we label SIS-1 and SIS-2. Including the more distant companion IC~4357 we find that NGC~5433 is the dominant member of a group with at least 5 members, spanning over 750~kpc in a filamentary structure. Optical and \ion{H}{1} distortions in NGC~5433 itself, and in particular the pathologies in the velocity fields of the better-resolved companions, indicate that interactions are occurring in the vicinity of NGC~5433, either among the galaxies or with hot intergalactic gas that may be present in the group. 

A variety of underlying mechanisms are consistent with the extraplanar feature morphologies in NGC~5433. The locations of the two peaks in F3 and their apparent correspondence with 20 cm radio continuum loops in a high-resolution map suggest that either a large nuclear outflow or a galactic fountain is at work. However, the active environment in NGC~5433 may be partly responsible for the extensions, either via direct impacts, tidal displacement of material from the midplane or by enhancing star formation in the disc.  We argue that environmental effects may play a role in both internally and externally generated high-latitude extensions in spiral galaxies.


\section*{Acknowledgments}
KS thanks Shami Chatterjee and Tara Chaves for useful comments and suggestions regarding data reduction and analysis. The observations have been financed (for JAI) by the 
Natural Sciences and Engineering Research Council of Canada.  
The NASA/IPAC Extragalactic Database
(NED) is operated by the Jet Propulsion Laboratory, California
Institute of Technology, under contract with the National Aeronautics and
Space Administration.  The Digitized Sky Survey was produced at the Space Telescope Science 
Institute under U.S. Government grant NAG W-2166.  
This publication makes use of data products from the Two Micron All Sky Survey, which is a joint project of the University of Massachusetts and the Infrared Processing and Analysis Center/California Institute of Technology, funded by the National Aeronautics and Space Administration and the National Science Foundation.
This research has made use of the MAPS Catalog of POSS I, which is supported by the National Aeronautics and Space Administration and the University of Minnesota. The APS databases can be accessed at http://aps.umn.edu/.

\appendix

\section[]{RFI Excision}
\label{rfi}

This appendix describes the techniques employed to remove strong radio frequency interference (RFI) at 1400 MHz from our VLA data, and our assessment of the impact of
this RFI excision on our results. Other data reduction details are given in \S\ref{obsred}.

The application of standard flux, phase and bandpass calibration routines to the data revealed undulations at constant declination 
in the central channels of all our (independently obtained) datasets.  The culprit was the 1400 MHz band centre adopted for the observations: with this setup, correlated noise from the 7th harmonic of the VLA's 200 MHz local oscillator at L-band contaminated the shortest east-west baselines in the central channels, for which the expected fringe rate is zero (Bagri 1996). For each calibrated dataset, this RFI was excised by {\it a)} removing baselines with east-west projections close to the array centre {\em in all channels}, and {\it b)} clipping contaminated data on amplitude in the visibility domain from the {\em infected central channels only}. We apply $a)$ to all channels because existing {\sevensize AIPS} software did not allow channel selection, and because no flux is lost from the initially RFI-free channels with this approach (see below).

The data were edited iteratively until a combination of parameters in methods {\it a)} and {\it b)} yielded the lowest RMS level $\sigma$ in the central channels while eliminating the RFI at that level. In the end, baselines with east-west projections within 20 metres of the array centre were flagged in each cube with the {\sevensize AIPS} task {\sevensize UVNOU}, and the optimum central channel clip levels for method {\it b)} were found to be 1.5 and 0.8 Jy for the C and D configuration datasets, respectively. For both methods $a)$ and $b)$ combined, 9 per cent of the baselines were removed in the central channel ($V=4352\,\rm{km}\,\rm{s}^{-1}$) for the C data, and in the D data 7 per cent of the baselines were excised at this velocity. For all datasets, we measure a $\,\sim10$ per cent increase in $\sigma$ at $V=4352\,\rm{km}\,\rm{s}^{-1}$ relative to the corresponding datacube mean. The same editing scheme was also applied in the channels adjacent to $V=4352\,\rm{km}\,\rm{s}^{-1}$ that exhibited low-level striping; in all cases less than 3 per cent of baselines were removed, and less than a 5\% increase in $\sigma$ was measured in the resulting maps at these velocities.

We verified that editing the RFI-free channels using method {\it a)} had no effect on the \ion{H}{1} emission detected at those frequencies by imaging calibrated cubes in which the editing had and had not been applied: we find no coherent structures in a difference map of those cubes in these channels.  To test for a systematic decrease of the \ion{H}{1} flux in NGC~5433 in the central channels due to RFI excision with methods {\it a)} and {\it b)}, we removed the same baselines in adjacent channels (that didn't exhibit RFI) and imaged them. We examined the difference between the images obtained without this additional editing and those for which it had been applied, in the vicinity of NGC~5433. In these residual maps, there is emission near NGC~5433 with flux at a level $\sim\sigma$ of the input cubes at $V=4363\,\rm{km}\,\rm{s}^{-1}$ and $V=4352\,\rm{km}\,\rm{s}^{-1}$ in the C-configuration data. This implies that the excision of RFI has resulted in a loss of flux in NGC~5433 at these velocities. We note that this exercise is strictly valid only if the \ion{H}{1} content in the channels used in the test is identical to that in the central channels. At the resolution of our maps, the distribution of \ion{H}{1} is similar near the systemic velocity of the galaxy, as is the total flux (see Fig. \ref{fig2}). It is therefore plausible that the residual map excess does indeed represent the extent of flux loss in NGC~5433. 

Because the potential flux loss is on the order of the map noise we do not correct for missing flux in any of our maps, nor do we expect our conclusions drawn from an examination of HI moment maps to change. For global parameter computations requiring \ion{H}{1} flux estimates of NGC~5433 in these central channels, we associate an additional, {\it systematic error} with the measured values, at twice the difference map peak flux in the vicinity of NGC~5433. This systematic error is represented by the dotted portion of the total error bars on these channels in the global profile of Fig.~\ref{fig2}. We have repeated this exercise for the detected companions of NGC~5433 with emission in those channels in the C+D data. We find no evidence for any flux loss, likely due to their smaller extent and distance from the phase centre.

We conclude that removing unwanted 1400 MHz interference from our data raises $\sigma$ at $V=4352\,\rm{km}\,\rm{s}^{-1}$ by $\sim$10 per cent relative to each datacube mean, and that in the C configuration data the central 2 channels may bias the measured flux for NGC~5433 by an amount of order $\sigma$ (see Fig.~\ref{fig2}). All other measurements, as well as our main conclusions, are unaffected by the RFI excision applied.  

\bsp

\label{lastpage}

\end{document}